\documentclass[twocolumn]{aastex631}

\newcommand\forecaster{\texttt{Forecaster}}
\newcommand\spock{\texttt{SPOCK}}

\date{April 1, 2185}

\shorttitle{Systems Alliance Planetary Survey}
\shortauthors{Obertas and Paradise}
\graphicspath{{./}{figures/}}

\begin{document}

\title{Preliminary Analysis of Planetary Characteristics, Dynamics, and Climates from the Systems Alliance Planetary Survey Catalogue}

\author{Alysa Obertas}
\email{obertas@astro.utoronto.ca}
\affiliation{David A. Dunlap Department of Astronomy \& Astrophysics, University of Toronto \\
50 St. George Street, Toronto, ON, M5S 3H4, Canada}
\affiliation{Canadian Institute for Theoretical Astrophysics \\
60 St. George Street, Toronto, ON, M5S 3H8, Canada}

\author{Adiv Paradise}
\email{paradise@astro.utoronto.ca}
\affiliation{David A. Dunlap Department of Astronomy \& Astrophysics, University of Toronto \\
50 St. George Street, Toronto, ON, M5S 3H4, Canada}

\begin{abstract}

Just a few decades after the discovery of the Charon Relay, and the ensuing First Contact War, relatively little is known about the population of planets linked by the Prothean mass relays. Understanding the nature of these systems and how they may differ from the broader population of planetary systems in our galaxy is key to both continued human habitation across the broader Galaxy, as well as to our understanding of the Prothean civilization. What factors motivated their choices of planetary systems? Characterizing these systems allows us to peer into Prothean society and culture, and make inferences about the preferences that drove their expansion throughout the Galaxy. In this study, we undertake a broad analysis of the systems recorded in the Systems Alliance Planetary Survey, examining their dynamical stability, orbital properties, and the climates of the inhabited worlds. We find that the Alliance data is inconsistent with both a modern understanding of planetary system dynamics, as well as with our understanding of Earth-like climate dynamics. We suggest this is due in part to security-related data obfuscation by the Alliance, and in part due to the real preferences of the Protheans.
\end{abstract}

\section{Introduction} \label{sec:intro}

The Systems Alliance has recently made their Planetary Survey available to the public. The survey began in 2182 and took 3 years to complete. It is the first in-depth survey conducted by humans of the stars and planets located in clusters accessible by Mass Relays. Not much is understood about the Mass Relays, but the prevailing hypothesis is that they were developed by the (now extinct) Prothean race, whose civilisations lie in ruins across the galaxy \citep{Tsoni2171}. Indeed, little is known about the Protheans at all. Consequently, analysing these planetary systems gives us a rich opportunity to probe why the Protheans wanted to access these worlds and provide us with means to better understand this lost race.

This survey also provides us with the means to examine in detail the systems that have been explored and colonised by the peoples of the galaxy. Human colonisation has been a prime focus of the Systems Alliance since 2152 and there are more than X human colonies across the galaxy. The safety and longevity of these worlds is crucial to protect our current colonies, in addition to those which have been identified as candidates for future colonisation efforts. 

In this preliminary analysis of the data provided in the SAPS catalogue, we explore the dynamics of multi-planet systems and the climates of select worlds in the catalogue. We present an overview of the SAPS catalogue in Section~\ref{sec:survey-overview}. We discuss our methods for our dynamics and climate analyses in Section~\ref{sec:methods} and present our results in Section~\ref{sec:results}. We present some discussion of our results (and the survey itself) in Section~\ref{sec:discussion} and then end with our conclusions in Section~\ref{sec:conclusions}

\vspace{10mm}

\section{Catalogue Overview} \label{sec:survey-overview}

The Systems Alliance Planetary Survey (SAPS) contains data on 148 stars and 580 planets. The targets of this survey are all in clusters accessible by Mass Relay and consequently, this is not an unbiased survey. The data contained in the catalogue is publicly available on the extranet \footnote{\url{https://masseffect.fandom.com/wiki/Planets}}.

\begin{figure}
\plotone{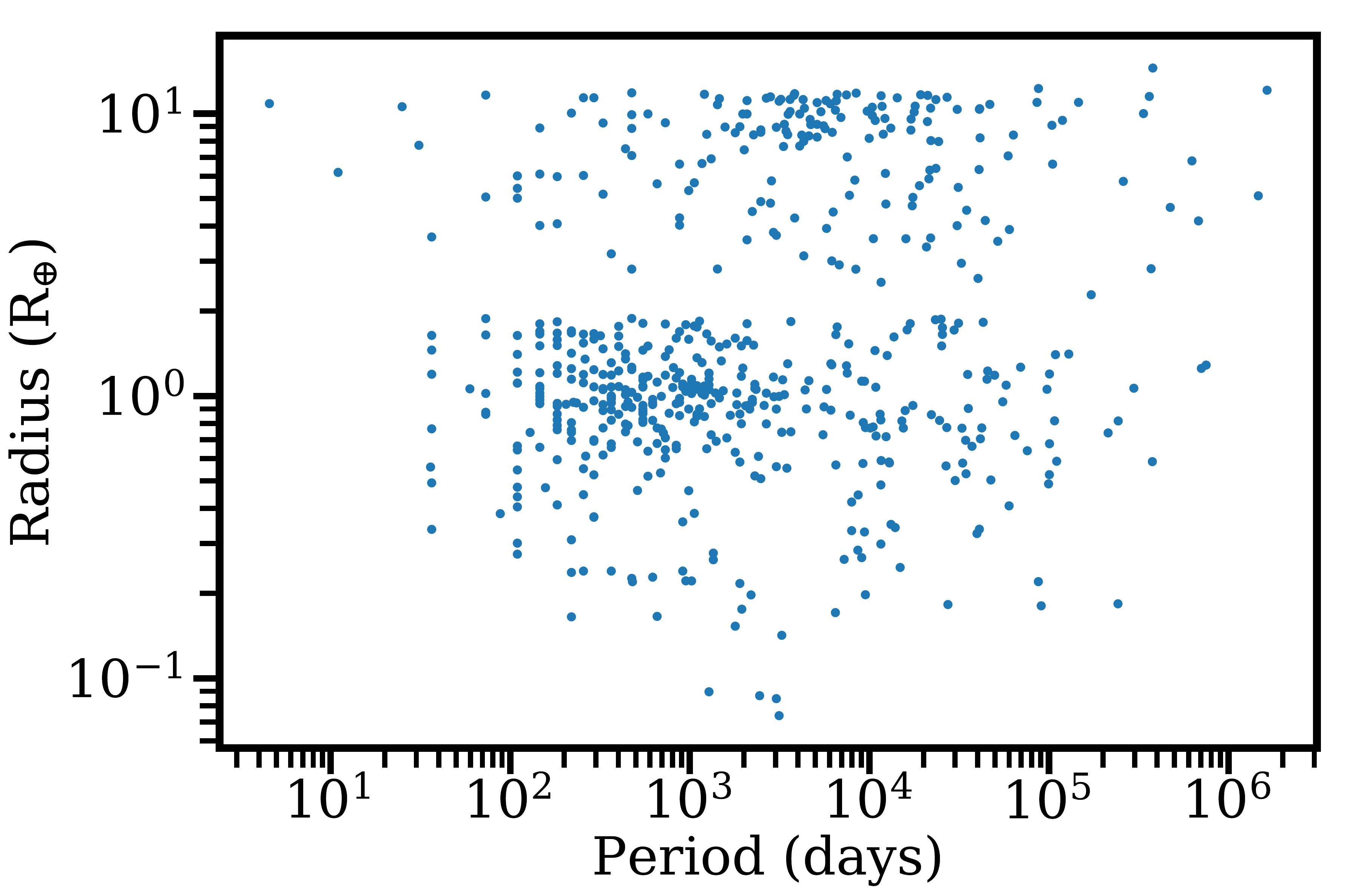}
\caption{The periods and radii of planets in the SAPS catalogue. All planets with both a radius and period are included. \label{fig:period-vs-radius}}
\end{figure}

Figure~\ref{fig:period-vs-radius} shows the periods and radii of planets in the SAPS catalogue. Planets are clustered around $R \sim R_{\oplus}$ and $P \sim 400$ days. Additionally, there is a distinct gap at $R \sim 2R_{\oplus}$.

\begin{figure}
\plotone{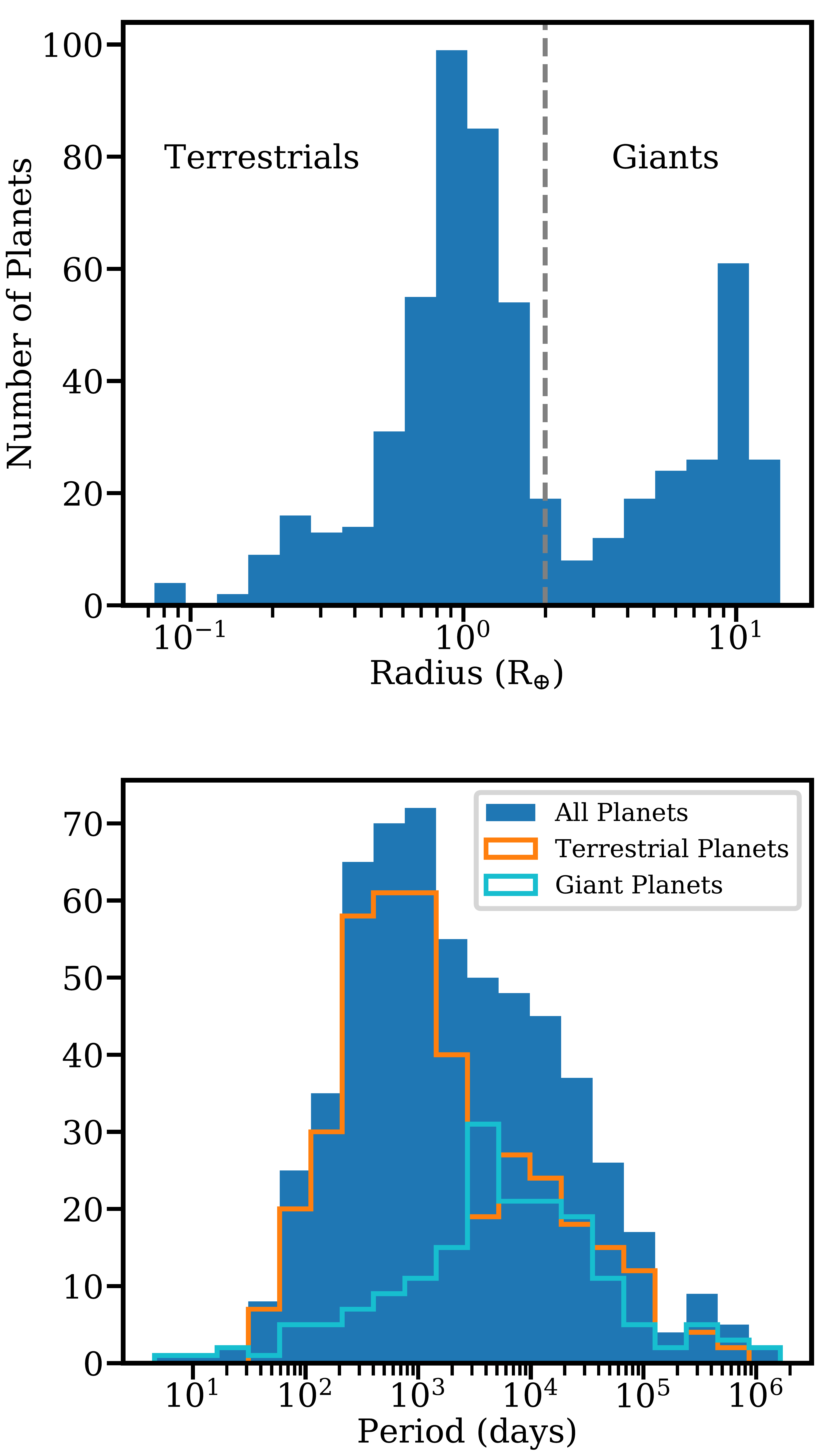}
\caption{Histograms showing the distribution of planet radii (left) and periods (right). The right plot also shows the separate distributions of periods for terrestrial planets ($R < 2 R_{\oplus}$, orange outline) and giant planets ($R \geq 2 R_{\oplus}$, cyan outline). Planets were binned according to the logarithm of their radius and period. \label{fig:period-radius-hist}}
\end{figure}

These features are more prominent in Figure~\ref{fig:period-radius-hist}, which shows the distributions of the periods and radii. There are two separate distributions for the planet radii, which represent terrestrial and giant planets. The terrestrial distribution peaks at $R \sim R_{\oplus}$ and the giant distribution peaks at $R \sim 10R_{\oplus}$. The terrestrial distribution sharply declines up to $R \sim 2 R_{\oplus}$. Note that the SAPS catalogue includes dwarf planets (i.e. anything large enough to be spherical).

The periods have a skewed distribution with a peak around $10^3$ days. The number of planets sharply increases between $P \ sim 100$ and $P \sim 1000$ days, with a tail extending up to $10^6$ days. Examining the period distributions separately for terrestrials and giants, it's clear that the peak around $10^3$ days is from the terrestrial population. The giants peak at $P \sim 3000$ days, although there are still a large proportion of giants at $P \sim 10^4$ days. 

The system multiplicities are given in Table~\ref{tab:multiplicity}. The second column (labelled Catalogue) shows the number of systems per multiplicity included in the SAPS catalogue. The third column (labelled Dynamics) shows the number of systems per multiplicity used in the analysis of system dynamics (see Section~\ref{subsec:method-dynamics} for details of which systems were included). There are 10 single-planet systems and 135 multi-planet systems. The mean multiplicity is 4.

\begin{deluxetable}{ccc}
\tablenum{1}
\tablecaption{\label{tab:multiplicity}Planet multiplicities of systems in the SAPS catalogue. The second column shows the number of systems in the catalogue and the third column shows the number of systems used for the dynamics analysis (see Section~\ref{subsec:method-dynamics} for a description of how systems were filtered).}
\tablewidth{1.0\columnwidth}
\tablehead{
\colhead{Multiplicity} & \colhead{Catalogue} & \colhead{Dynamics}
}
\startdata
$N=1$ & 10 & 0 \\
$N=2$ & 11 & 10 \\
$N=3$ & 24 & 21 \\
$N=4$ & 43 & 38 \\
$N=5$ & 43 & 33 \\
$N=6$ & 11 & 10 \\
$N=7$ & 2 & 2 \\
$N=8$ & 0 & 0 \\
$N=9$ & 1 & 1 \\
\enddata
\end{deluxetable}

\section{Methods} \label{sec:methods}

\subsection{System Dynamics} \label{subsec:method-dynamics}

We explored the dynamics of SAPS multi-planet systems in two separate categories: those with two planets and those with at least three planets (which we refer to as higher multiplicity systems). For the two-planet systems, the spacing was compared to the two-planet stability criterion \citep{gla93}. For the higher multiplicity systems, their stability was analysed using \spock~ \citep{tam20}.

To begin our analysis, we removed systems which did not have semimajor axes reported for its planets. This left 10 two-planet systems and 105 higher multiplicity systems (see Table~\ref{tab:multiplicity} for a breakdown of counts by multiplicity). 

Next, we calculated the stellar mass corresponding to each planet's semimajor axis and period (according to Kepler's 3rd law). Curiously, the stellar mass associated with each planet was not always consistent across a system (see discussion). As this is physically impossible, we implemented a method to obtain a single stellar mass for each system. First, we calculated the mean stellar mass and the standard deviation for each system. If the ratio of the standard deviation to the mean stellar mass was less than $10^{-3}$, we used the mean stellar mass as the single value for the system. Otherwise, we used only the stellar masses within one standard deviation of the mean stellar mass to calculate a new mean value (i.e. we removed the outliers) and used this as the single value for the system.

Now that we had a single stellar mass corresponding to each system, we re-calculated the semimajor axes based on each planet's orbital period. We assumed that the orbital periods reported by SAPS were correct, as it is substantially more straight-forward to calculate a planet's orbital period to high accuracy and precision. 

\subsubsection{Two-Planet Systems}

Examining the stability of the two-planet systems is as simple as comparing the spacing ($\Delta$) between planets to the critical spacing ($\Delta_C$) \citep{gla93}.

The spacing of planets ($\Delta$) is in terms of the mutual Hill radius ($R_{H,m}$), defined in terms of the semimajor axes of the inner and outer ($a_1$, $a_2$) planets and their planet-to-star mass ratios ($\mu_1$, $\mu_2$).

\begin{equation} \label{eq:hill-radius}
    R_{H,m} = \left(\frac{\mu_1 + \mu_2}{3}\right)\left(\frac{a_1+a_2}{2}\right)
\end{equation}

\begin{equation}
    \Delta = \frac{a_2 - a_1}{R_{H,m}}
\end{equation}

We use the form of critical spacing as given by \cite{obe17}, defined as

\begin{equation}
    \Delta_C = \frac{2\sqrt{3}}{1 + 2\sqrt{3}{X}}
\end{equation}

Here, $X$ is defined as

\begin{equation}
    X = \frac{1}{2}\left(\frac{\mu_1 + \mu_2}{3}\right)^{1/3}
\end{equation}

We used \forecaster~\citep{che17} to obtain masses for planets without masses reported in the SAPS catalogue. This was done by generating 1000 masses according to the planet's radius and then taking the median. 

\subsubsection{Higher Multiplicity Systems}

Despite the hundreds of years that astronomers have been working on problems in celestial mechanics, there are still no simple criteria to evaluate the stability of systems with three or more planets. Instead of analytic solutions, computational methods are necessary for exploring the stability of higher multiplicity systems in the SAPS catalogue.

We used \spock~\citep{tam20} to perform our analysis, as it is still the gold standard in this field. \spock~ is a machine learning tool which provides a probability of stability for $10^9$ orbits. We ran 5000 iterations of each system with \spock, varying the unknown planet masses and orbital elements with each iteration. For planets without masses reported in the SAPS catalogue, we used \forecaster~to generate a mass based on the planet's radius. The eccentricities and inclinations of planets were drawn from a Rayleigh distribution with parameters $\sigma_e=0.01$ and $\sigma_i=0.5^{\circ}$. The angles ($\omega, \Omega, f$) were drawn uniformly on the circle.

As this study is a preliminary analysis of the population as a whole, we did not take into consideration specifics of individual systems. An example is the system of Vernio: Promavess and Sotera are expected to have a collision within the next three years. This is clearly an unstable system which has already had orbit-altering close encounters.

\subsection{Climate} \label{sec:methods-climate}

Limited public data exists on the climates and atmospheres even of colonized planets, due in part to the Alliance Colony Security Act of 2158 (hereafter ACSA-2158). This poses additional challenges to understanding Prothean motivations and interests in certain planetary systems. We can however attempt to infer additional information about the climates of colonized worlds near mass effect relays by simulating their climates using 3D general circulation models, or GCMs. GCMs have been used to study a range of planets, including Earth \citep{Way2017}, Mars \citep{marsgcm}, Venus \citep{venusgcm}, and numerous exoplanets \citep[e.g.][]{Komacek2019,westeros}. In this study, we use the ExoPlaSim GCM, an intermediate-complexity GCM developed from Earth climate models, and adapted for exoplanets \citep{Paradise2020,exoplasim}. ExoPlaSim uses a spectral dynamical core to solve the primitive fluid equations, and includes models for cloud formation, precipitation, sea ice, and oceanic thermal properties through a 50-meter mixed-layer slab ocean \citep{Fraedrich2005}. We use ExoPlaSim in its T21 resolution, corresponding to 32 latitudes and 64 longitudes. 

\begin{figure}
\plotone{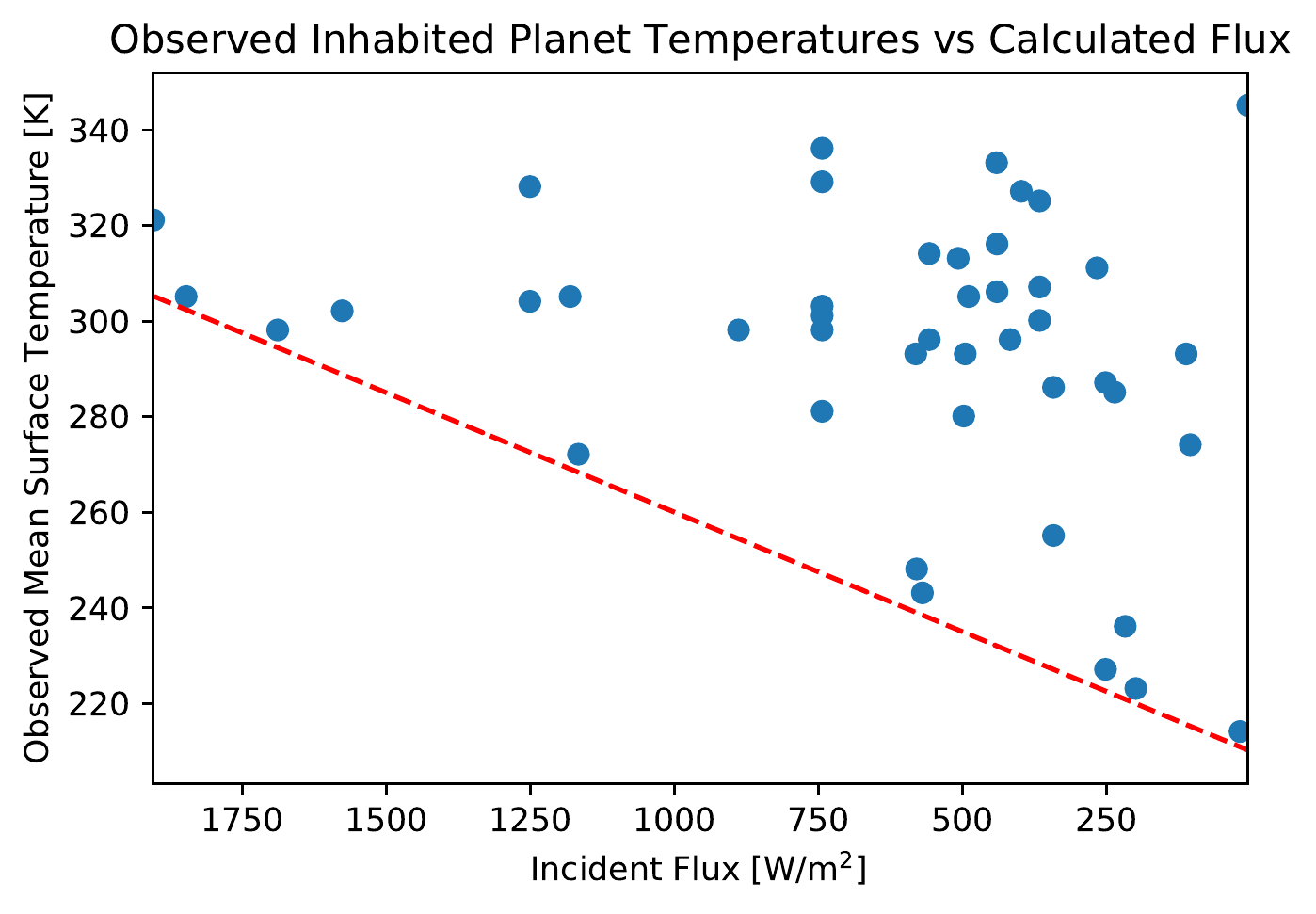}
\caption{Observed mean surface temperatures vs calculated incident fluxes. We assume a mass-luminosity relationship of $L\propto M^{3.5}$, which is appropriate for most main-sequence stars. We use the semimajor axes and stellar masses derived in \autoref{subsec:method-dynamics}. Note that mean surface temperatures are mostly uncorrelated to incident flux, more planets are found at lower fluxes, and the minimum temperature observed on a colonized world at a given flux increases linearly with flux. \label{fig:realtempsfluxes}}
\end{figure}

A particular challenge of studying the climates of colonized worlds in the era of ACSA-2158 is that GCMs require the incident stellar flux as an input to the model, which requires both stellar luminosity and the planet's semimajor axis. As noted above, however, Alliance data on planet semimajor axes is unreliable. If we compare observed mean surface temperatures to incident fluxes computed from the semimajor axes and stellar masses derived in \autoref{subsec:method-dynamics} (assuming $L\propto M^{3.5}$, as shown in \autoref{fig:realtempsfluxes}, we find that most colonies are found at low incident fluxes (for reference, Earth receives 1367 W/m$^2$). Colony temperatures are also mostly uncorrelated with incident flux, with the exception of the coldest colonies found at a given flux, whose temperatures increase linearly with flux. This is consistent with the expectation that cold worlds should simply be rarer at higher fluxes \citep{Kopparapu2014}. The finding that there are numerous warm planets at low insolations is however at odds with our understanding of the carbon-silicate cycle, by which planets at low incident fluxes should experience limit cycles and spend most of their time in fully-glaciated snowball states \citep{Paradise2017}. This could either be explained by colonized worlds in systems with mass relays having significantly higher levels of volcanic activity and outgassing than that found on Earth, and therefore greenhouse gas abundances significantly different from that found on Earth, or by deliberate obfuscation of Alliance data. 

To demonstrate this, we model a selection of colonized worlds, focusing primarily on those with mean surface temperatures between 200 and 350 K (due to model limitations), and surface pressures between 0.1 and 10 bars, also due to model limitations \citep{Paradise2021}. We run each model with a timestep of either 10 or 15 minutes, using the SAPS-provided planetary radii, surface gravities, rotation rates, and surface pressures, for 50 years each. This is generally not enough to reach complete energy-balance equilibrium, but is enough to provide a qualitative picture of the likely climate, leaving top-of-atmosphere errors in energy budget of only a few W/m$^2$. To facilitate comparison to our Earth-based expectations of habitability, we assume N$_2$-dominated atmospheres, with trace greenhouse gases. To avoid biasing our results by assumptions made in our calculation of semimajor axes and stellar masses, we initially assume incident flux is unconstrained, and for each planet, run several iterations at different fluxes. We scale the range of fluxes considered to the observed mean surface temperature, so that we capture lower fluxes for colder planets, expecting that incident flux will be the dominant factor in overall climate, as it is on Earth. We also consider three different CO$_2$ abundances: 0 ppm, representing climates with low volcanism in which H$_2$O is the only major greenhouse gas, 400 $\mu$bar, analogous to Earth in the early 21st century, and 10 mbar, analogous to Earth during the warmest parts of its recent geologic history. 

As ozone is highly-dependent on atmospheric chemistry and photochemistry, and primarily affects stratospheric temperatures, we ignore ozone for the purposes of this study. We also simplify the incident stellar spectra, assigning each planet's host star a simple blackbody temperature based on its broad spectral classification (M, K, G, F, or A). These blackbody temperatures are 3500 K, 4500 K, 5800 K, 7000 K, and 8500 K respectively. ExoPlaSim accounts for the shape of the incident spectrum through the partitioning of two shortwave radiation bands ($<0.75 \mu$m and $>0.75 \mu$m), the albedos of the various surface types such as water and sea ice, and through the strength of Rayleigh scattering.

As a limiting case, we assume each planet is an aquaplanet, with uniform sea surface. This is obviously inappropriate, especially for land-dominated planets such as Tuchanka, but as ocean has a very low albedo (0.069 under a Solar spectrum), and water is a powerful greenhouse gas, this is likely to increase simulated temperatures for moderate incident fluxes. We do note however that the presence of land can increase local temperatures to the point of habitability for planets with widespread glaciation \citep{Paradise2019}, so there are likely to be more partially-habitable planets at low fluxes than our results will show (and AP in fact notes this with first-hand experience, due to a lengthy post-baccalaureate internship on Noveria). Our methods will primarily be useful for demonstrating the minimum fluxes at which we might expect to find truly Earth-like planets which are globally-temperate, for the bulk planetary properties known for each colony. Our results are given in \autoref{sec:results-climate}.
\section{Results} \label{sec:results}

\subsection{Dynamics}

A summary of results for two-planet systems is shown in Table~\ref{tab:two-planet}. Two systems (Chandrasekhar and Nariph) are more closely-spaced than the critical value and the remaining eight systems are more widely-spaced than the critical value. Both of these systems contain planets without masses reported in the SAPS catalogue and were generated using \forecaster, however. A maximum combined mass of the two planets ($M_{max}$) can be calculated by setting the spacing at the critical value,

\begin{equation} \label{eq:max-mass}
    M_{max} = M_1 + M_2 = 3M_{\star} \left(\frac{a_2-a_1}{4\sqrt{3}a_1}\right)^3
\end{equation}

\begin{deluxetable*}{lccccc}
\tablenum{2}
\tablecaption{\label{tab:two-planet}Summary of stability for two-planet systems. The critical spacing ($\Delta_C$), spacing ($\Delta$), and the inner and outer planet masses ($M_1$ and $M_2$). Spacing is in units of mutual Hill radius (eq.~\ref{eq:hill-radius}). Systems with $\Delta < \Delta_C$ also include a maximum combined mass of the two planets (eq.~\ref{eq:max-mass}).}
\tablewidth{0pt}
\tablehead{
\colhead{System} & \colhead{\hspace{.75cm}$\Delta_C$}\hspace{.75cm} & \colhead{\hspace{.75cm}$\Delta$}\hspace{.75cm} & \colhead{\hspace{.5cm}$M_1$ ($M_{\oplus}$)}\hspace{.5cm} &
\colhead{\hspace{.5cm}$M_2$ ($M_{\oplus}$)}\hspace{.5cm} & \colhead{$M_{max}$ ($M_{\oplus}$)}
}
\startdata
Chandrasekhar & 2.51 & 1.79 & 3000.70 & 27.92 & 899.92 \\
Solveig & 3.29 & 23.23 & 0.384 & 27.06 & \\
Caestus & 3.39 & 70.43 & 1.929 & 0.054 & \\
Kalabsha & 3.02 & 8.28 & 1.134 & 595.11 & \\
Nariph & 2.56 & 1.66 & 2595.17 & 10.42 & 561.98 \\
Alpha Draconis & 2.85 & 6.76 & 1.141 & 2677.33 & \\
Phi Clio & 3.27 & 20.96 & 31.29 & 0.668 & \\
Decoris & 3.39 & 44.38 & 0.937 & 1.295 & \\
Chomos & 2.77 & 7.24 & 1982.24 & 0.517 & \\
Loropi & 2.91 & 5.56 & 2705.09 & 0.078 & \\
\enddata
\end{deluxetable*}

For Chandrasekhar and Nariph the values of $M_{max}$ are 899.92$M_{\oplus}$ and 561.98$M_{\oplus}$, respectively. Both of these systems contain an inner giant ($R_1$ = 9.98$R_{\oplus}$ and 10.05$R_{\oplus}$) and an outer ice giant (i.e. the inner planet dominates the combined mass). The values of $M_{max}$ for both systems are consistent with giant planets, so the most likely explanation is that \forecaster~ is providing masses that are too large for these specific planets.

\begin{figure}
\plotone{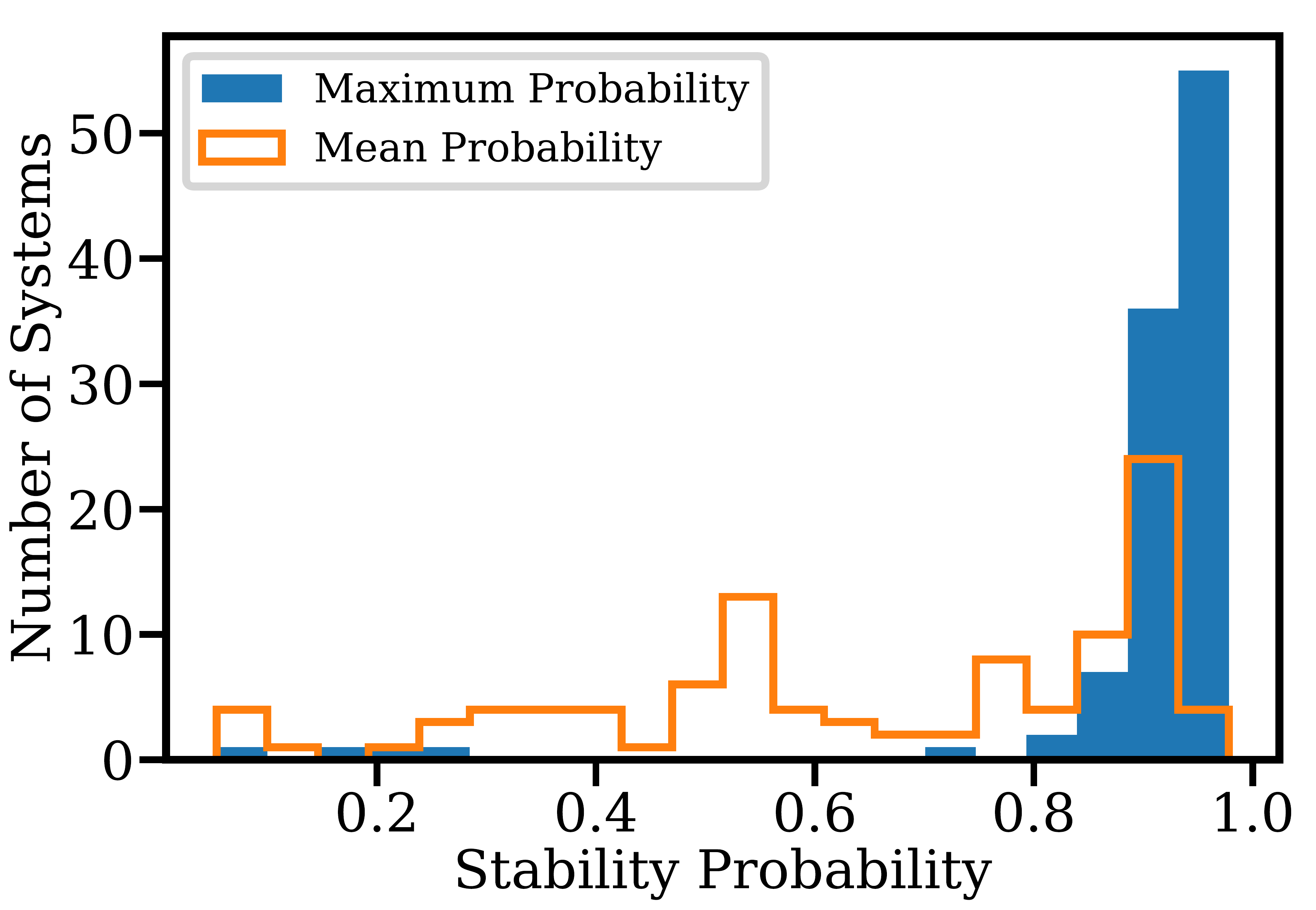}
\caption{Histograms of the maximum and mean stability probability for systems with at least three planets. Each system was run 5000 times with \spock, varying the unknown planet and orbit properties (i.e. those not included in the SAPS catalogue). A system's maximum stability probability is the maximum of the 5000 iterations (and similarly, the mean for the mean probability). \label{fig:prob-hist}}
\end{figure}

To analyse the stability of systems with at least three planets, we selected the configuration which had the highest probability of survival for $10^9$ orbits, as determined by \spock. Figure~\ref{fig:prob-hist} shows a histogram of the maximum and mean probabilities for these systems. If not specified, when we discuss a system's probability we are referring to its maximum probability.

There are four systems with low probabilities: Kriseroi, Mil, Skepsis, and Micah. In Kriseroi, Neidus and Theonax are the source of this low probability. The SAPS catalogue lists Neidus and Theonax with semimajor axes of 0.1AU and 0.18AU, but both with periods of 0.1 years. This is not physically possible, suggesting an error in the catalogue listing. In Mil, Chalkhos and Selvos are in a binary configuration, which was not implemented in the stability analysis. In Skepsis, Pauling's data in the SAPS catalogue is identical to that of Crick (i.e. this is a catalogue error). For Micah, it is not clear why the probabilities are all very low, and further analysis would be necessary (e.g. integrating the system). Neglecting these four systems, the mean probability is 0.925. The lowest probability is 0.745 (Dranek) and the highest is 0.978 (Ploitari).

The mean probabilities are also of interest, as this indicates whether the maximum probability was an outlier, or if the system has a high stability probability regardless of the unknown parameters used. Again neglecting the same four systems, the mean of the mean probability is 0.663. The lowest (mean) probability is 0.0367 (Han) and the highest is 0.957 (Ploitari). 

We also calculated the standard deviation of each system's 5000 probabilities. The mean value is 0.183, the smallest is 0.00575 (Talava), and the highest is 0.418 (Malgus). 

It is not surprising that Ploitari has both the highest mean and maximum probabilities: it is a four-planet system with relatively low-mass terrestrials ($M_1=0.006 M_{\oplus},~M_2=1.283 M_{\oplus},~M_3=0.596 M_{\oplus},~M_4=0.175 M_{\oplus}$). The standard deviation of this system's probabilities is 0.0217, which is fairly small.

\begin{figure}
\plotone{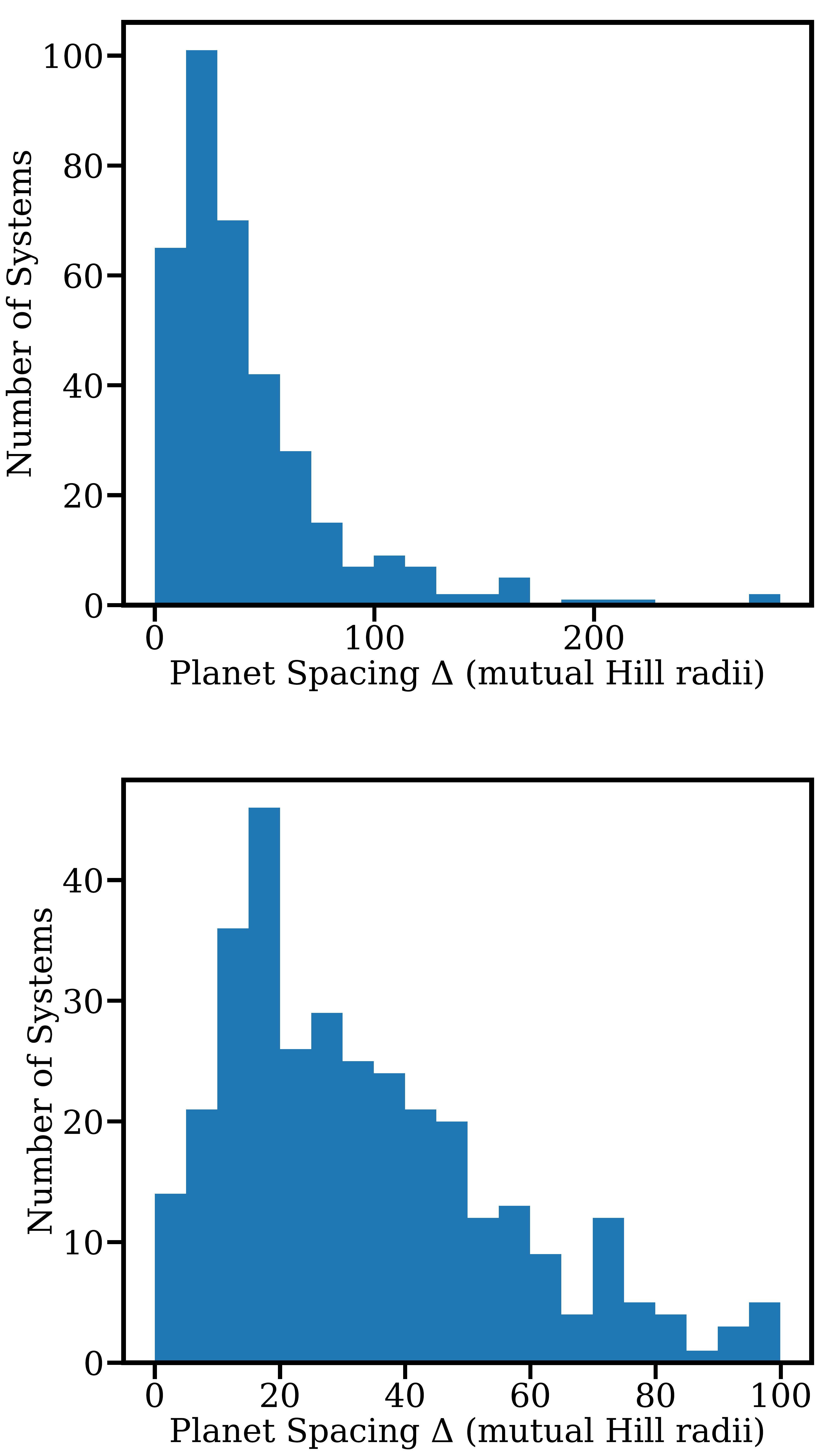}
\caption{Histograms of the spacing between adjacent pairs of planets ($\Delta$), in mutual Hill radii, for systems with at least three planets. The top plot shows the distribution for all values of $\Delta$. The bottom plot shows the distribution for $\Delta < 100$. \label{fig:delta-hist}}
\end{figure}

The spacing between adjacent pairs of planets ($\Delta$) can also be examined for these systems. For planets without masses listed in the SAPS catalogue, we used the mass (generated by \forecaster~ based on the planet's radius) corresponding to the iteration with the maximum probability. The distribution of $\Delta$ is shown in Figure~\ref{fig:delta-hist}. There is a range of spacing, but the distribution peaks at $\Delta \sim 20$. Neglecting pairs with identical semimajor axes, the mean spacing is 42.79, the smallest spacing is 2.24 (Malgus), and the largest spacing is 284.82 (Farinata).

The results of our stability analysis are shown in Table~\ref{tab:high-mult}. This includes the mean probability ($p_{\mathrm{mean}}$), maximum probability ($p_{\mathrm{ax}}$), standard deviation of the probability ($\sigma_p$), and mean spacing ($\overline{\Delta}$) for all systems examined. We calculated the mean spacing by taking the mean value of each planet's mass (unless listed in the SAPS catalogue), calculating each adjacent pair's spacing, and then taking the mean of those values.

\startlongtable
\begin{deluxetable*}{lcccc}
\tablenum{3}
\tablecaption{\label{tab:high-mult}Summary of stability for higher multiplicity systems. The columns show the mean probability, maximum probability, probability standard deviation, and mean spacing for each system. The mean spacing is calculated by taking the mean value of a planet's mass (unless listed in the SAPS catalogue), calculating each adjacent pair's spacing, and then taking the mean of those values.}
\tablewidth{0pt}
\tablehead{
\colhead{System} & \colhead{\hspace{.75cm}$p_{\mathrm{mean}}$}\hspace{.75cm} & \colhead{\hspace{.75cm}$p_{\mathrm{max}}$}\hspace{.75cm} & \colhead{\hspace{.75cm}$\sigma_p$}\hspace{.75cm} &
\colhead{\hspace{.75cm}$\overline{\Delta}$}\hspace{.75cm}
}
\startdata
Gorgon  & 0.482 & 0.925 & 0.310 & 31.857 \\ 
Hydra  & 0.867 & 0.943 & 0.135 & 29.295 \\ 
Phoenix  & 0.075 & 0.902 & 0.114 & 37.820 \\ 
Macedon  & 0.759 & 0.885 & 0.124 & 33.263 \\ 
Hercules  & 0.853 & 0.935 & 0.110 & 79.468 \\ 
Theseus  & 0.492 & 0.923 & 0.369 & 29.264 \\ 
Utopia  & 0.767 & 0.944 & 0.288 & 71.418 \\ 
Asgard  & 0.633 & 0.801 & 0.098 & 20.050 \\ 
Han  & 0.037 & 0.820 & 0.123 & 15.356 \\ 
Ming  & 0.913 & 0.945 & 0.019 & 95.881 \\ 
Antaeus  & 0.362 & 0.919 & 0.394 & 47.206 \\ 
Cacus  & 0.902 & 0.944 & 0.043 & 59.232 \\ 
Dis  & 0.923 & 0.947 & 0.011 & 40.013 \\ 
Farinata  & 0.908 & 0.940 & 0.015 & 225.170 \\ 
Plutus  & 0.737 & 0.939 & 0.176 & 101.443 \\ 
Century  & 0.537 & 0.971 & 0.389 & 51.699 \\ 
Schwarzschild  & 0.407 & 0.935 & 0.363 & 13.726 \\ 
Verr  & 0.375 & 0.925 & 0.387 & 15.153 \\ 
Fortuna  & 0.906 & 0.935 & 0.015 & 47.936 \\ 
Pax  & 0.932 & 0.948 & 0.009 & 91.564 \\ 
Newton  & 0.619 & 0.948 & 0.374 & 42.187 \\ 
Sol  & 0.851 & 0.900 & 0.028 & 24.127 \\ 
Boltzmann  & 0.574 & 0.907 & 0.338 & 37.632 \\ 
Acheron  & 0.912 & 0.933 & 0.015 & 93.166 \\ 
Aysur  & 0.507 & 0.910 & 0.348 & 49.978 \\ 
Balor  & 0.917 & 0.937 & 0.016 & 112.187 \\ 
Talava  & 0.930 & 0.943 & 0.006 & 60.851 \\ 
Yakawa  & 0.592 & 0.917 & 0.246 & 17.225 \\ 
Lusarn  & 0.916 & 0.937 & 0.012 & 53.370 \\ 
Ondeste  & 0.946 & 0.967 & 0.009 & 39.002 \\ 
Tasale  & 0.319 & 0.841 & 0.283 & 28.596 \\ 
Zelene  & 0.400 & 0.968 & 0.388 & 25.121 \\ 
Amun  & 0.519 & 0.916 & 0.321 & 27.259 \\ 
Imir  & 0.777 & 0.941 & 0.094 & 27.283 \\ 
Malgus  & 0.529 & 0.960 & 0.418 & 30.321 \\ 
Relic  & 0.240 & 0.883 & 0.295 & 15.279 \\ 
Dholen  & 0.767 & 0.942 & 0.309 & 3.374 \\ 
Hekate  & 0.883 & 0.930 & 0.023 & 34.414 \\ 
Hoplos  & 0.918 & 0.947 & 0.029 & 77.182 \\ 
Pamyat  & 0.894 & 0.934 & 0.017 & 57.172 \\ 
Faryar  & 0.138 & 0.923 & 0.193 & 39.840 \\ 
Osun  & 0.523 & 0.929 & 0.272 & 26.333 \\ 
Ploitari  & 0.957 & 0.978 & 0.022 & 53.527 \\ 
Sowilo  & 0.373 & 0.849 & 0.210 & 28.602 \\ 
Aquila  & 0.203 & 0.861 & 0.266 & 18.991 \\ 
Elysta  & 0.789 & 0.909 & 0.125 & 31.624 \\ 
Faia  & 0.832 & 0.935 & 0.115 & 50.375 \\ 
Aralakh  & 0.082 & 0.904 & 0.205 & 69.524 \\ 
Dranek  & 0.377 & 0.745 & 0.128 & 33.293 \\ 
Nith  & 0.918 & 0.941 & 0.009 & 122.460 \\ 
Arrae  & 0.556 & 0.913 & 0.373 & 39.581 \\ 
Fortis  & 0.930 & 0.943 & 0.007 & 58.056 \\ 
Dakka  & 0.658 & 0.926 & 0.338 & 26.338 \\ 
Amada  & 0.834 & 0.919 & 0.032 & 68.836 \\ 
Batalla  & 0.919 & 0.943 & 0.011 & 26.697 \\ 
Fathar  & 0.938 & 0.967 & 0.006 & 51.030 \\ 
Kairavamori  & 0.910 & 0.936 & 0.012 & 46.955 \\ 
Sahrabarik  & 0.551 & 0.931 & 0.347 & 13.837 \\ 
Dirada  & 0.259 & 0.902 & 0.303 & 58.483 \\ 
Kriseroi  & 0.024 & 0.054 & 0.012 & 18.955 \\ 
Satent  & 0.874 & 0.921 & 0.017 & 49.202 \\ 
Zaherin  & 0.554 & 0.858 & 0.123 & 34.308 \\ 
Enoch  & 0.292 & 0.923 & 0.279 & 31.081 \\ 
Iera  & 0.802 & 0.917 & 0.177 & 33.421 \\ 
Lenal  & 0.823 & 0.944 & 0.225 & 39.160 \\ 
Mil  & 0.072 & 0.191 & 0.050 & 36.774 \\ 
Psi Tophet  & 0.593 & 0.936 & 0.304 & 34.550 \\ 
Skepsis  & 0.076 & 0.198 & 0.043 & 43.496 \\ 
Tassrah  & 0.929 & 0.943 & 0.007 & 44.541 \\ 
Typhon  & 0.593 & 0.930 & 0.361 & 45.114 \\ 
Kyzil  & 0.522 & 0.938 & 0.412 & 50.677 \\ 
Thal  & 0.893 & 0.943 & 0.045 & 45.395 \\ 
Urla Rast  & 0.874 & 0.950 & 0.205 & 31.673 \\ 
Xe Cha  & 0.514 & 0.969 & 0.301 & 26.277 \\ 
Micah  & 0.029 & 0.268 & 0.046 & 19.828 \\ 
Bahak  & 0.277 & 0.908 & 0.327 & 50.954 \\ 
Aru  & 0.775 & 0.922 & 0.255 & 20.114 \\ 
Esori  & 0.544 & 0.891 & 0.125 & 47.912 \\ 
Nura  & 0.529 & 0.927 & 0.371 & 24.236 \\ 
Satu Arrd  & 0.938 & 0.959 & 0.012 & 111.332 \\ 
Pranas  & 0.688 & 0.931 & 0.317 & 37.032 \\ 
Castellus  & 0.874 & 0.936 & 0.054 & 62.916 \\ 
Trebia  & 0.444 & 0.856 & 0.257 & 35.784 \\ 
Arcturus  & 0.865 & 0.925 & 0.053 & 30.970 \\ 
Euler  & 0.893 & 0.944 & 0.031 & 43.813 \\ 
Ialessa  & 0.784 & 0.937 & 0.162 & 38.301 \\ 
Orisoni  & 0.358 & 0.923 & 0.379 & 28.495 \\ 
Parnitha  & 0.419 & 0.903 & 0.336 & 34.095 \\ 
Tomaros  & 0.611 & 0.925 & 0.353 & 17.289 \\ 
Vernio  & 0.290 & 0.938 & 0.327 & 29.143 \\ 
Harsa  & 0.736 & 0.943 & 0.305 & 42.839 \\ 
Indris  & 0.481 & 0.933 & 0.395 & 32.869 \\ 
Untrel  & 0.908 & 0.938 & 0.016 & 42.797 \\ 
Vular  & 0.922 & 0.945 & 0.021 & 31.755 \\ 
Kallini  & 0.530 & 0.949 & 0.210 & 41.277 \\ 
Mesana  & 0.309 & 0.902 & 0.336 & 25.024 \\ 
Pelion  & 0.930 & 0.977 & 0.037 & 48.327 \\ 
Maskim Xul  & 0.928 & 0.947 & 0.015 & 29.843 \\ 
Mulla Xul  & 0.505 & 0.943 & 0.358 & 37.949 \\ 
Tikkun  & 0.888 & 0.921 & 0.011 & 48.909 \\ 
Vetus  & 0.561 & 0.904 & 0.288 & 40.859 \\ 
Kypladon  & 0.883 & 0.946 & 0.052 & 32.644 \\ 
Nahuala  & 0.766 & 0.934 & 0.198 & 32.055 \\ 
Phontes  & 0.867 & 0.939 & 0.063 & 30.675 \\ 
Teyolia  & 0.545 & 0.936 & 0.369 & 42.661 \\ 
\enddata
\end{deluxetable*}

\subsection{Climate} \label{sec:results-climate}

\begin{figure}[h]
\plotone{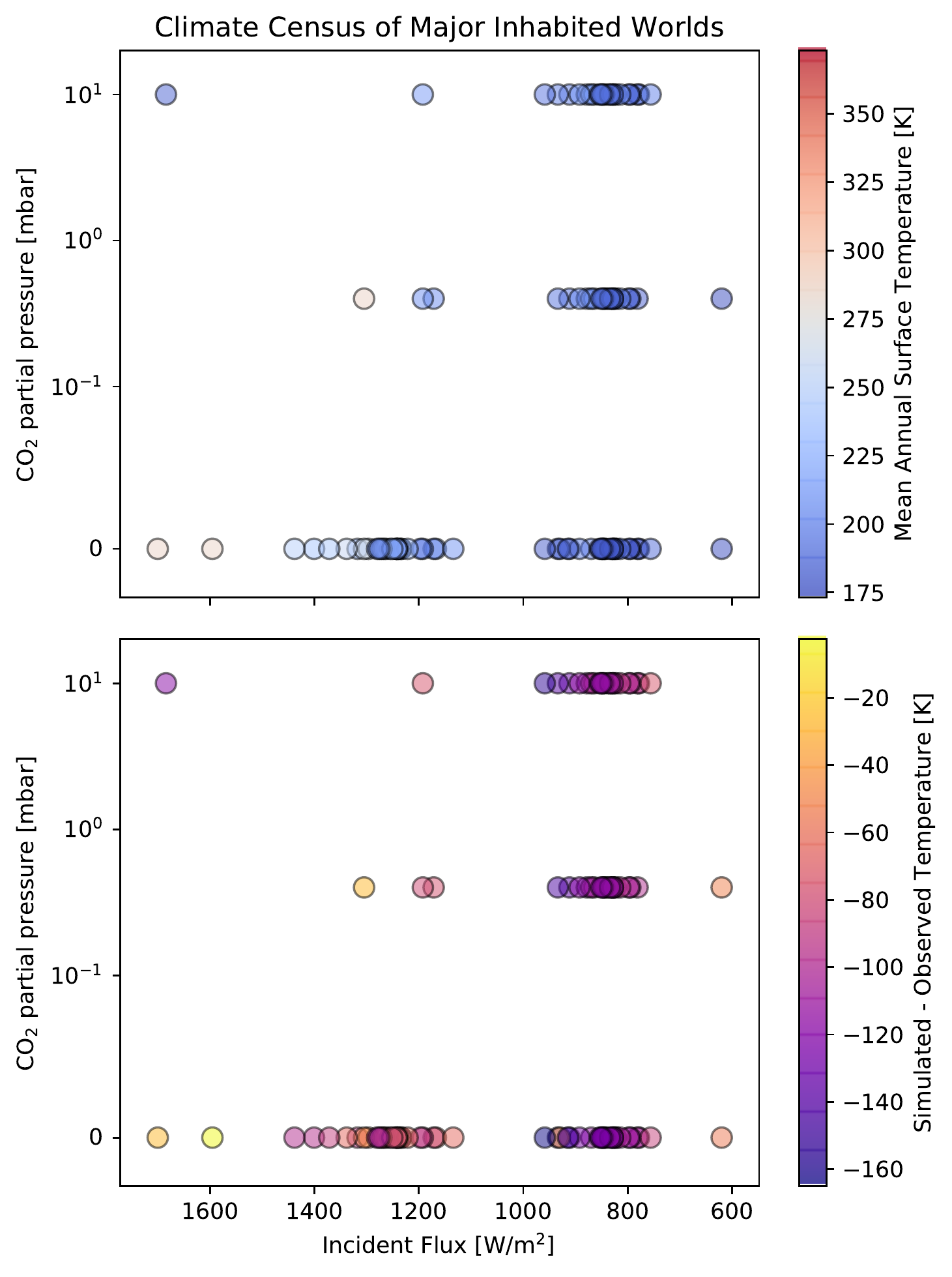}
\caption{Mean annual surface temperatures modeled with ExoPlaSim for 102 models, representing 28 colonized and/or inhabited worlds. We also give the difference between modeled temperatures and real temperatures. Note that almost all our models are far colder than real observed temperatures, with the exception of a few models at higher fluxes. This is consistent with expectations that Earth analogues should freeze over at relatively high incident fluxes \citep{Paradise2017}. \label{fig:modeltempsfluxes}}
\end{figure}
\begin{figure}
\plotone{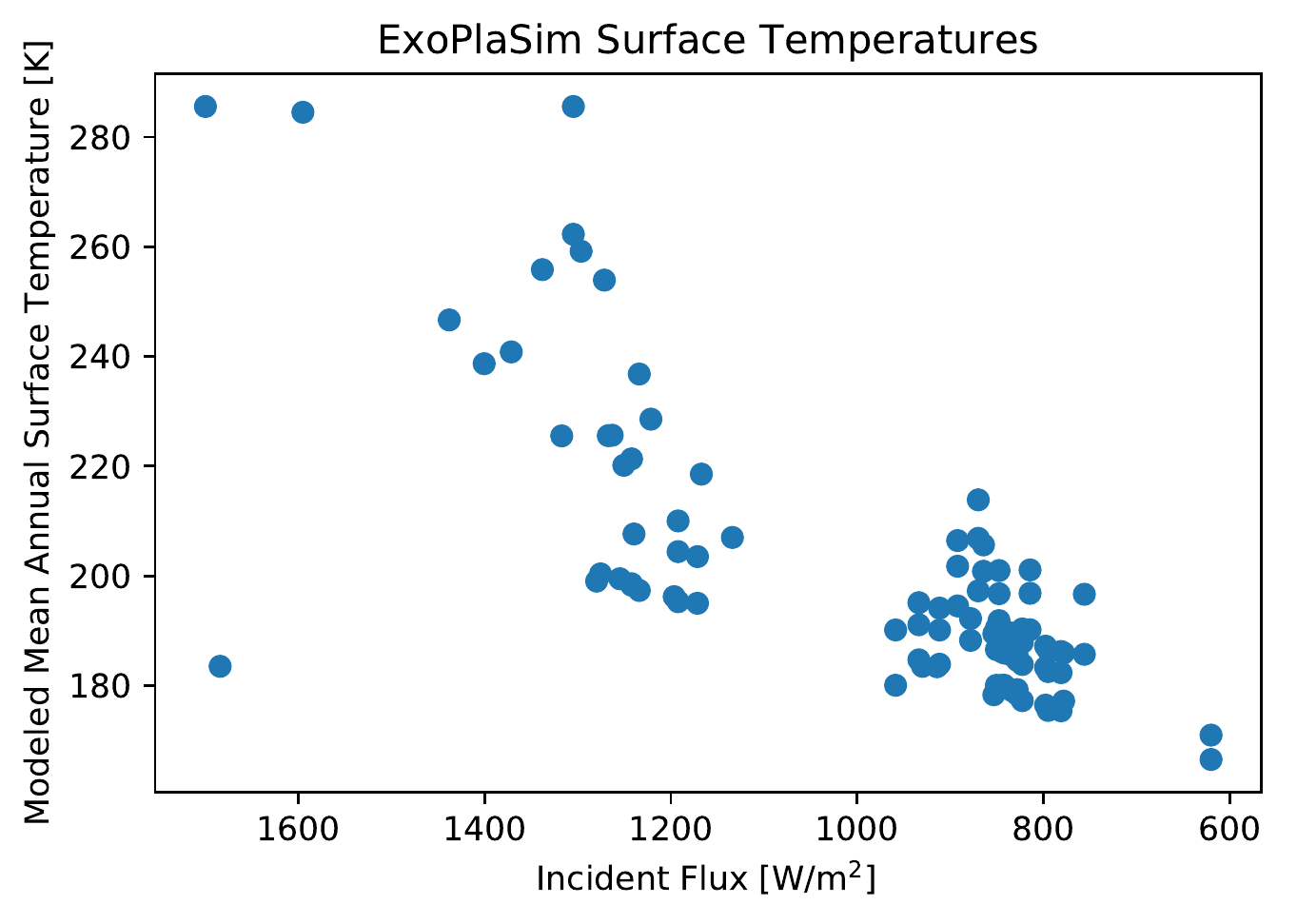}
\caption{Modeled mean surface temperature vs incident flux for the models simulated in \autoref{sec:methods-climate}. Note that in our population of models, there is a strong correlation between incident flux and mean surface temperature, unlike the observed distribution shown in \autoref{fig:realtempsfluxes}. It may therefore be the case that Earth represents the kind of climate found along the minimum-temperature boundary shown as the dashed red line in \autoref{fig:realtempsfluxes}, rather than the typical habitable planet that may have been preferred by the Protheans. \label{fig:modeledtemps}}
\end{figure}

\begin{figure*}[ht!]
\plotone{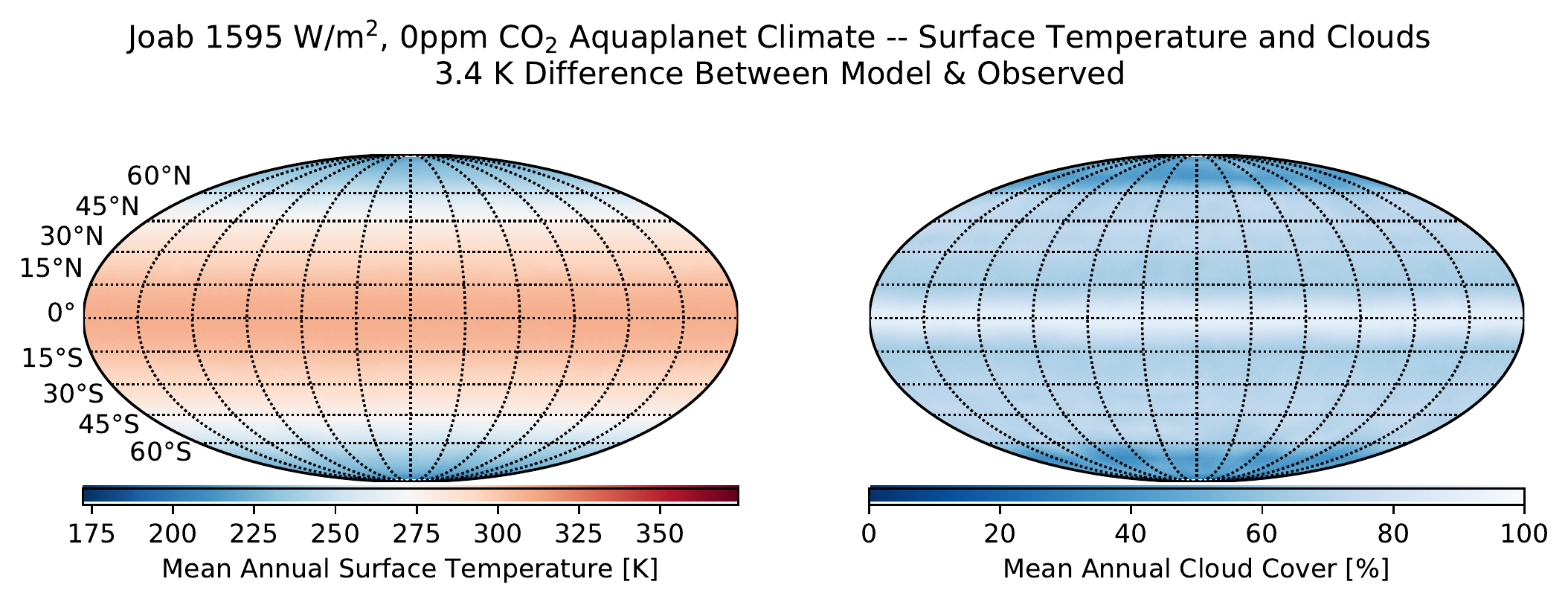}
\caption{An example of a modeled climate that agrees relatively closely with observed real temperatures (Joab, in the Enoch system). Note however that based on the stellar mass and semimajor axis we calculate based on SAPS data, Joab receives 287.15 W/M$^2$, not the 1595 W/m$^2$ ExoPlaSim needed to produce these surface temperatures with an Earth-like atmosphere and low volcanism. This suggests that either Joab is much closer to its star than the 2.3 AU suggested by SAPS data, or it has an extremely non-Earth-like atmosphere. \label{fig:joab}}
\end{figure*}

The mean annual surface temperatures of the climates we modeled are shown in \autoref{fig:modeltempsfluxes}. We find that at the incident fluxes deemed likely by the semimajor axes we computed in \autoref{subsec:method-dynamics} and the observed surface temperatures reported by colonists, almost all of our models are very cold, very unlike the warm temperatures shown in \autoref{fig:realtempsfluxes}. In fact, the temperature-flux distribution we find, as shown in \autoref{fig:modeledtemps}, looks much more like the minimum-temperature population cutoff shown as the dashed red line in \autoref{fig:realtempsfluxes}. We do find some models that agree well in terms of mean surface temperature, such as Joab, in the Enoch system. Surface temperature and cloud cover for our model of Joab are shown in \autoref{fig:joab}. However, ExoPlaSim required an incident flux of nearly 1600 W/m$^2$ to produce these temperatures with an atmosphere consistent with low volcanism, which is more than 5 times the incident flux Joab receives according to our computed semimajor axis and stellar mass. Joab is a small super-Earth at 1.3 M$_\oplus$ and 1.05 R$_\oplus$, but it has a thicker atmosphere of 2.18 atm. ExoPlaSim can account for this, but what it cannot account for is the dust in the atmosphere created by the massive orbital bombardment Joab sustained within the last few kyr. We had expected this dust to primarily cool through increased scattering, but it could be that the dust is also contributing to the greenhouse effect. 

\section{Discussion} \label{sec:discussion}

\subsection{Dynamics} \label{subsec:dynamics-discussion}

Overall, the multi-planet systems in the SAPS catalogue are quite stable. Nearly all systems have a maximum probability of at least 0.745. Even considering all configurations tested, most systems still had mean probabilities of at least 0.663 with a typical standard deviation of 0.183. This suggests that when assembling the Mass Relay network, the Protheans likely considered the stability of systems in a cluster.

Ideally, complete N-body integrations would be run for each system to obtain a more detailed and thorough understanding of their individual dynamics and stability. We opted not to pursue this in our preliminary analysis, and instead to focus on overall trends in the stability of systems. Additionally, this would require much more time to account for the planet and orbit characteristics not included in the SAPS catalogue.

We suspect that these parameters were measured during the survey, but have not been released to the public. Furthermore, we believe that the data included in the catalogue has been intentionally altered. We base this opinion on the problems with stellar masses described in Section~\ref{subsec:method-dynamics}.

We calculated the relative error of the semimajor axes according to

\begin{equation}
    \sigma_a = \frac{|a_{\mathrm{SAPS}} - a_{\mathrm{calculated}}|}{a_{\mathrm{SAPS}}}
\end{equation}

Here, $a_{\mathrm{SAPS}}$ is the semimajor axis listed in the SAPS catalogue and $a_{\mathrm{calculated}}$ is the semimajor axis we calculated. The mean error was 3.59\%, the median error was 0.435\%, and the maximum error was 155\%. 12.8\% (63 of the 492 which had semimajor axes in the SAPS catalogue and one that we could calculate) of planets had an error of at least 5\% and 37 planets (7.52\% of the 492) had an error of at least 10\%. Given the survey's ability to make precise and accurate measurements of stellar and planetary masses, we are shocked by these discrepancies. 

\subsection{Climate}

While we are shocked by the discrepancies in semimajor axis reported by SAPS, and note that this represents a continued hindrance to planetary science by ACSA-2158, and thus to the free movement and sovereignty of the Galaxy's inhabitants, we also note that these errors are not sufficient to explain the discrepancies in climate between our simulations and the observed surface temperatures  of the inhabited worlds in our sample. In particular, they cannot account for the distinctly non-astrophysical correlation shown in \autoref{fig:realtempsfluxes}. We do not expect volcanism to vary strongly with incident flux, so we suggest that this discrepancy may in fact reflect a Prothean preference that informed the placement and maintenance of the mass relays. 

We therefore propose that in addition to ensuring they only invested resources in stable systems, Protheans had a distinct preference for systems with habitable worlds at low incident fluxes, bearing atmospheres rich in greenhouse gases. While ACSA-2158 limits the information we have on the atmospheric compositions of the inhabited worlds, we can say with some certainty that most are at least breathable by humans. Their greenhouse gas abundances must therefore be dominated by nontoxic gases. We propose that the Prothean home world may have had abundant volcanism, with geochemistry and biochemistry that could support abundant CH$_4$, CO$_2$, and water, with perhaps other greenhouse gases that are not found in any significant quantities on Earth. The fact that they frequented worlds whose atmospheres are breathable further suggests that $N_2$ and $O_2$ may have also been major components of their home world. We urge further study using higher-complexity GCMs and a diverse range of atmospheric compositions, using our computed semimajor axes as inputs for estimating incident flux. 

An alternative explanation may be that the Prothean home world was in fact more Earth-like than most planets in our sample, but they visited worlds with non-Earth-like atmospheres for other practical considerations, such as a need to avoid damaging stellar flares. The Protheans are known to have maintained an expansive galactic empire, with a particular eye towards stability and permanence \citep{Tsoni2171}. It may be that in their study of potential systems, they preferentially chose systems whose habitable worlds received less stellar flux, and therefore were at less risk of damaging flares that could destroy crucial infrastructure.

\section{Conclusions} \label{sec:conclusions}

The Systems Alliance Planetary Survey (SAPS) catalogue was recently made available to the public, containing stellar and planetary parameters for clusters accessible by the Mass Relay Network. We conducted a preliminary analysis of the population, dynamics, and climate of systems and worlds in this survey.

Our main findings can be summarised as

\begin{itemize}
    \item Planets are either terrestrial ($R < 2 R_{\oplus}$) or giant ($R > 2 R_{\oplus}$)
    \item Terrestrial planets are clustered around periods of 400 days and giants are clustered around periods of 3000 days
    \item Multi-planet systems show a high degree of stability. Two-planet systems are more widely-spaced than the critical limit. Higher multiplicity systems show high probability of stability, with a typical maximum probability of 92.5\%.
    \item Inhabited planets linked by the mass relays have generally temperate climates, even at low incident fluxes that should render Earth-like climates uninhabitable, indicating robust greenhouse gas abundances that suggest a low-flux, high-volcanism Prothean home world. 
\end{itemize}

We also note that there are errors and inconsistencies in the SAPS catalogue, which we attribute to intentional obfuscation. This is possibly due to the Alliance Colony Security Act of 2158, but we argue that this legislation is out of date. This is especially true now that humanity is part of the Citadel Council and establishing its place in the galaxy.

Based on our analysis, we present the following recommendations

\begin{itemize}
    \item Precise and accurate data be provided for stellar and planetary properties
    \item High-precision orbital data be provided for planets, especially those in multi-planet systems
    \item A comprehensive program of atmospheric retrievals undertaken by independent science teams to characterize the greenhouse gas abundances of inhabited planets
    \item The Systems Alliance re-visit its legislation regarding the release of scientific information to the public
\end{itemize}

\begin{acknowledgments}
We wish to acknowledge this land on which the University of Toronto operates. For thousands of years it has been the traditional land of the Huron-Wendat, the Seneca, and most recently, the Mississaugas of the Credit River. Today, this meeting place is still the home to many Indigenous people from across Turtle Island and we are grateful to have the opportunity to work on this land. We also wish to acknowledge the impact that high-performance computing has on the environment and on Indigenous peoples, both through its energy cost, and through the destructive impact of the precious metal mining that is necessary for computer components.

Given the content of this paper, we also encourage people to question and discuss the language and ideas surrounding human colonisation beyond our planet. This is typically framed from a Eurocentric perspective that mirrors the history of colonisation on Earth. At present space exploration is dominated by national governments and militaries, with some private corporations acting under the aegis of national governments. This work further demonstrates the risks of concentrating space exploration and knowledge of space in institutions that may not always act democratically or in everyone's best interests. 

We also would like to thank everyone at BioWare who has worked on the Mass Effect series and for creating such a phenomenal series of games. We are so excited for the upcoming Legendary Edition and the new title currently in progress.

Additionally, we want to thank all of the fans who have entered information on the Mass Effect Wiki (\url{https://masseffect.fandom.com/wiki/Mass_Effect_Wiki}) over the years. This paper would not have been possible without easy access to this data and information.

Alysa would like to thank Norm Murray and Dan Tamayo for understanding why she spent the last week working on this paper, instead of her real paper that she should have been working on. Adiv would like to thank his collaborators for their assistance testing the Exoplasim model, and their patience, and acknowledges financial support from the Province of Ontario through the Ontario Graduate Scholarship. Access to computing infrastructure was provided by the Canadian Institute for Theoretical Astrophysics.

We also acknowledge gratefully the gracious endorsement of Commander Shepard, who assured us this was her favourite study on the planetary systems linked by the Prothean mass relay networks.
\end{acknowledgments}

\software{numpy \citep{numpy},
          \forecaster \citep{che17},
          \spock \citep{tam20}
          Rebound \citep{rei12}
          astropy \citep{2013A&A...558A..33A,2018AJ....156..123A},  
          Cloudy \citep{2013RMxAA..49..137F}, 
          Source Extractor \citep{1996A&AS..117..393B}
          ExoPlaSim \citep{exoplasim}
          }


\bibliography{bibliography}{}
\bibliographystyle{aasjournal}



\end{document}